\documentclass[letterpaper,twocolumn,10pt]{article}

\usepackage{silence}
\usepackage{article-style}

\hypersetup{hypertexnames=false}
\usepackage{xurl}

\usepackage{tikz}
\usetikzlibrary{shapes.geometric, arrows.meta, positioning, shadows, backgrounds, fit}
\usepackage{tcolorbox}
\tcbuselibrary{breakable, skins}
\usepackage{pgfplots}
\pgfplotsset{compat=1.18}

\usepackage{amsmath}
\usepackage{amssymb}
\usepackage{bbm}

\usepackage{booktabs}
\usepackage{tabularx}
\usepackage{multirow}
\usepackage{array}

\usepackage{float}
\usepackage{stfloats}
\usepackage{caption}

\usepackage{algorithm}
\usepackage{algpseudocode}

\usepackage{listings}
\usepackage{orcidlink}
\usepackage{soul}
\usepackage[scale=0.85]{sourcecodepro}
\usepackage{calc}

\DeclareRobustCommand{\redact}{\mbox{\rule{1.5em}{1.2ex}}}
\newcommand{\redactshort}{\mbox{\rule{0.8em}{1.2ex}}}

\definecolor{slateaccent}{RGB}{215, 222, 228}
\definecolor{slate}{RGB}{246, 248, 250}
\definecolor{offblack}{RGB}{34, 39, 46}
\definecolor{green}{RGB}{58, 174, 75}
\definecolor{lightgreen}{RGB}{230, 244, 232}
\definecolor{lightgreenaccent}{RGB}{170, 230, 175}
\definecolor{red}{RGB}{248, 81, 73}
\definecolor{lightred}{RGB}{255, 235, 233}
\definecolor{lightredaccent}{RGB}{255, 180, 180}
\definecolor{gold}{RGB}{210, 153, 35}
\definecolor{lightgold}{RGB}{254, 247, 224}
\definecolor{blue}{RGB}{12, 105, 218}
\definecolor{lightblue}{RGB}{230, 241, 255}
\definecolor{orange}{RGB}{232, 83, 0}
\definecolor{lightorange}{RGB}{252, 241, 235}
\definecolor{purple}{RGB}{130, 80, 223}
\definecolor{lightpurple}{RGB}{245, 240, 255}
\definecolor{cyan}{RGB}{8, 145, 178}
\definecolor{lightcyan}{RGB}{224, 247, 250}
\definecolor{pink}{RGB}{219, 97, 162}
\definecolor{lightpink}{RGB}{255, 239, 247}
\definecolor{gray}{RGB}{101, 109, 118}
\definecolor{lightgray}{RGB}{230, 232, 235}

\begin{document}

\date{}

\title{\Large \bf AEGIS: White-Box Attack Path Generation using LLMs and\\
  Training Effectiveness Evaluation for Large-Scale Cyber Defence Exercises}

\author{
{\rm Ivan K.\ Tung\,\orcidlink{0000-0001-9454-1905}, Shi Yu Xiang\,\orcidlink{0009-0004-4870-4290}, Alex Chien\,\orcidlink{0009-0001-9727-2509}, Liu Wenkai\,\orcidlink{0009-0005-1953-7523}, Lawrence Zheng\,\orcidlink{0009-0005-9623-3347}}\\
\\
{Cyber Defence Test and Evaluation Centre (CyTEC),}\\
{The Digital and Intelligence Service (DIS), Singapore Armed Forces}
}

\maketitle

\begin{abstract}
	Creating attack paths for cyber defence exercises requires substantial expert effort. Existing automation requires vulnerability graphs or exploit sets curated in advance, limiting where it can be applied. We present AEGIS, a system that generates attack paths using LLMs, white-box access, and Monte Carlo Tree Search over real exploit execution. LLM-based search discovers exploits dynamically without pre-existing vulnerability graphs, while white-box access enables validating exploits in isolation before committing to attack paths. Evaluation at CIDeX 2025, a large-scale exercise spanning 46 IT hosts, showed that AEGIS-generated paths are comparable to human-authored scenarios across four dimensions of training experience (perceived learning, engagement, believability, challenge). Results were measured with a validated questionnaire extensible to general simulation-based training. By automating exploit chain discovery and validation, AEGIS reduces scenario development from months to days, shifting expert effort from technical validation to scenario design.
\end{abstract}

\section{Introduction}

Cyber defence exercises (CDX) are critical for providing realistic scenarios to train cyber defenders to detect and recover from attacks~\cite{karjalainen_comprehensive_2020}. However, designing realistic scenarios requires immense effort~\cite{russo_cyber_2023, katsantonis_cyber_2023}. Creating a credible attack path requires experts to identify exploits, configure command-and-control frameworks, and chain tactics, techniques, and procedures (TTP) into a coherent story. For example, during the Critical Infrastructure Defence Exercise (CIDeX) 2025~\cite{ministry_of_defence_singapore_ai-enabled_2025}, a large-scale cyber defence exercise, crafting two attack paths for a three day scenario required up to 30 person-months of expert red-teamer effort.

\begin{tcolorbox}[colback=slate, colframe=slateaccent, boxrule=1pt]
	\textbf{Observation 1:} Current attack planning methods for cyber defence exercises are time-consuming.
\end{tcolorbox}

While automation tools exist, they are disjointed and insufficient for end-to-end exercise generation. Traditional vulnerability scanners such as Nessus~\cite{tenable_inc_nessus_nodate} or OpenVAS~\cite{greenbone_ag_openvas_nodate} rely on static databases and usually require manual verification of exploitability. Reinforcement learning (RL) approaches can optimise attack paths, but depend on pre-existing vulnerability graphs~\cite{mckinnel_systematic_2019}. Most LLM-based penetration testing tools avoid these dependencies, but evaluations on multi-host environments are limited to narrow scopes such as Active Directory~\cite{happe_can_2025} or constrained action sets~\cite{singer_feasibility_2025}, and their black-box nature cannot leverage the ground-truth network knowledge available to CDX designers.

\begin{tcolorbox}[colback=slate, colframe=slateaccent, boxrule=1pt]
	\textbf{Observation 2:} Existing approaches to automate attack path planning for cyber defence exercises are limited.
\end{tcolorbox}

In response to these observations, we introduce AEGIS (Automated Exercise Generation through Intelligent Simulation), an attack path generation system that uses LLMs and Monte Carlo Tree Search (MCTS) to assist red teams in crafting exercise scenarios. This paper makes the following contributions:

\medskip
\noindent \textbf{An automated white-box attack planning workflow that reduces man-effort from months to days.} Given full access to a network, we present a LLM-based workflow that decomposes attack path generation into stages, where ground-truth information at key stages enables errors to be caught before they compound. Such errors often contribute to compounding decision-making failures that would stall existing LLM tools.

\medskip
\noindent \textbf{An MCTS method for attack path mapping with real exploit execution.} As opposed to typical RL-based simulations, we introduce an MCTS method where actions are real exploit executions. To compensate for limited sampling due to real-time execution being orders of magnitude slower than simulated steps, AEGIS reduces the action space through partial exploit validation and guided search with LLM heuristics, without requiring offline training.

\medskip
\noindent \textbf{A statistically validated instrument to assess CDX training effectiveness.} We present a 12-item questionnaire grounded in pedagogical principles and validated with factor analysis, isolating four dimensions of training experience extensible to general simulation-based training.

\section{Related Work}
\label{sec:related-work}

\subsection{Automated Attack Planning}

Existing tools like MulVAL~\cite{ou_mulval_2005} generate attack graphs from pre-existing vulnerability data without validating exploitability. Formal planning enables principled reasoning over graphs: Sarraute et al.~\cite{sarraute_penetration_2013} demonstrate scalability to medium-sized networks, though real-world deployment remains limited. RL approaches typically frame penetration testing as a Markov Decision Process and train policies in simulators~\cite{schwartz_autonomous_2019}, but transfer poorly to real environments~\cite{nguyen_pengym_2025}. Planning-based approaches using MCTS have been applied to attack path analysis over pre-computed graphs~\cite{xie_improved_2018}, network security games~\cite{xue_nsgzero_2022}, and vulnerability-guided search~\cite{tu_vital_2024}. However, applying these to real exploit execution, where actions take minutes rather than milliseconds, remains unexplored.

\subsection{LLM-Based Penetration Testing}

LLM agents for penetration testing face two recurring architectural challenges: context loss causing repeated actions, and difficulty backtracking after failed exploits~\cite{deng_pentestgpt_2024, zhu_teams_2025}. Single-agent approaches address context loss using structured state: PentestGPT~\cite{deng_pentestgpt_2024} uses Pentesting Task Trees (PTT), while AutoPT~\cite{wu_autopt_2024} decomposes tasks into phases via a finite state machine. Multi-agent architectures like HPTSA~\cite{zhu_teams_2025} address backtracking by delegating recovery to a supervisory planner. Incalmo~\cite{singer_feasibility_2025} and Cochise~\cite{happe_can_2025} extend to multi-host attacks but remain constrained by manually curated action sets and the narrow scope of Active Directory respectively. Most other tools have only been evaluated on single hosts or flat topologies. These tools operate in black-box settings, discovering the network as they attack, but CDX designers control the infrastructure with full knowledge and access, which existing tools do not leverage.

\subsection{CDX Automation and Evaluation}

Prior CDX automation primarily addresses infrastructure or content; attack path generation remains limited. AiCEF~\cite{zacharis_aicef_2023} extracts scenario narratives from threat intelligence reports but does not produce executable attack paths. CALDERA~\cite{mitre_corporation_caldera_nodate} executes adversary emulation but from manually configured profiles. Lore~\cite{holm_lore_2023} is most similar: it uses RL to execute attack paths, but extending to new exploits requires manual integration. Open-ended generation without such manual curation remains unaddressed.

CDX scenario evaluation is challenging. Metrics like incident report accuracy can capture operational outcomes but cannot assess subjective scenario qualities such as realism. Post-exercise questionnaires are standard practice~\cite{seker_concept_2018}, but they are typically not validated. Lore's questionnaire~\cite{holm_realistic_2025}, the only such published CDX instrument, measures perceived learning, realism, and challenge, but uses single items for the first two constructs and lacks factor validation.

\section{Exploratory Study}
\label{sec:exploratory-study}

While LLM-based agents can search for and execute arbitrary exploits, traditional attack planning tools still require pre-curated exploits to generate attack paths. Before designing a new system, we evaluated whether existing LLM tools could fill this gap.

\begin{description}
	\item[RQ1] Can existing LLM tools create attack paths for a CDX?
\end{description}

\subsection{Testing Strategy}

\begin{figure}[hb]
	\centering
	\includegraphics[width=\columnwidth]{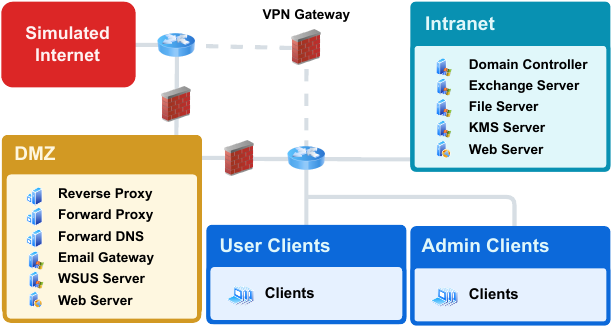}
	\caption{\label{fig:network-topology} Network topology.}
\end{figure}

Existing evaluations of LLM pentesting tools focus on single hosts~\cite{deng_pentestgpt_2024, zhu_teams_2025, fang_llm_2024, jiang_autopengpthighly_2024}, flat topologies~\cite{xu_autoattacker_2024}, narrow scopes such as Active Directory~\cite{happe_can_2025}, or simulated environments~\cite{genevey-metat_red_2023}. The only multi-host benchmark we identified, MHBench~\cite{singer_feasibility_2025}, was not publicly available at the time of evaluation. Unlike these prior evaluations, we tested on the CIDeX 2025 IT segment (Figure~\ref{fig:network-topology}), a realistic network from a large-scale CDX comprising 4 Linux servers, 7 Windows servers, 30 Windows clients, and 5 network appliances including a VPN gateway and Active Directory infrastructure. The network has a tree topology with four compartments: DMZ, Intranet, User Clients, and Admin Clients.

We evaluated two LLM tools. The first was Cochise~\cite{happe_can_2025}, a specialised open-source tool that addresses multi-host attacks and allows the LLM to run arbitrary commands. The second was OpenHands~\cite{wang_openhands_2025}, a general-purpose coding agent that was the top-performing open-source agent on TerminalBench~\cite{laude_institute_laude-instituteterminal-bench_2025} at the time of evaluation. Some modifications were necessary for evaluation on our network. These are listed in Appendix~\ref{app:modifications}.

We used both tools with Claude Sonnet 4.5 (\texttt{20250929}). We conducted three runs per tool, ranging from 2 to 5.5 hours each until infrastructure issues halted execution, all exceeding the 2-hour runtime used in Cochise's original evaluation~\cite{happe_can_2025}.

\subsection{Findings}

Both tools succeeded at reconnaissance, discovering hosts, services, and potential vulnerabilities. However, neither progressed to lateral movement (Table~\ref{tab:stage-results}).

\begin{table}[ht]
	\centering
	\small
	\begin{tabularx}{\columnwidth}{X|cc}
		\toprule
		\textbf{Stage} & \textbf{Cochise} & \textbf{OpenHands} \\
		\midrule
		Reconnaissance & \checkmark & \checkmark \\
		Initial Access & \texttimes & \checkmark \\
		Post-Exploitation & --- & \texttimes \\
		Lateral Movement & --- & --- \\
		\bottomrule
	\end{tabularx}
	\caption{Stages reached across runs of Cochise and OpenHands. \checkmark{} (achieved in all runs), \texttimes{} (failed in all runs), --- (never attempted).}
	\label{tab:stage-results}
\end{table}

Cochise identified CVE-2023-2745, Smart Slider 3 plugin vulnerabilities, and other installed WordPress plugins, but none were exploitable on the target. OpenHands achieved initial access in all runs via two Metasploit exploits (CVE-2023-6553, CVE-2024-2054), but encountered repeated reverse shell instability and could not progress to lateral movement.

Earlier studies found that even scaffolded LLM systems exhibit reasoning deficits such as context loss and false command generation~\cite{deng_pentestgpt_2024}. Our observations here differ: both tools could maintain context across multi-step reconnaissance. Failures occurred not at reasoning but at decision points requiring pentesting judgement (e.g.\ when to test exploits, how broadly to explore, when to prepare for risks), suggesting earlier limitations may no longer apply to current models.

\subsection{Comparison with Expert Behaviour}

To understand these failures, we analysed each tool's bottleneck: the attack chain stage where it failed to progress. Within each bottleneck, we identified decisions made by the LLM that diverged from what a skilled  human pentester with the same information would typically have done. We use expert judgement rather than formal frameworks (e.g.\ PTES, NIST 800-115) because such frameworks describe process phases, not the tactical decision-making observed here. Our evaluation uses high-controllability attribution~\cite{weiner_attributional_1985}: if a human would have succeeded, the failure is attributed to a controllable agent choice rather than an environmental constraint.

\textbf{Cochise at Initial Access.} Cochise's approach was limited. It rejected exploits based on version information rather than testing them empirically, despite version mismatches not always precluding possible exploitation. It also focused on low-yield information-disclosure vulnerabilities on the web server rather than enumerating other hosts with code execution opportunities in all runs. In our experience, skilled pentesters typically test quickly and move on when an approach yields diminishing returns.

\textbf{OpenHands at Post-Exploitation.} OpenHands obtained shell access but encountered session instability eight times in total across the three runs. Rather than upgrading to a stable shell, the agent performed the same exploit again each time a session died to reestablish the connection, then immediately resumed exploration. During post-exploitation, it scanned only a single internal subnet rather than common private ranges, limiting discovery of lateral movement targets. In our experience, skilled pentesters typically stabilise access first by upgrading the shell, then systematically enumerate internal networks before proceeding.

We characterise three patterns where the agent's decision-making varied from our determined typical expert behaviour:

\smallskip
\noindent \textbf{Test vs Assume.} Skilled pentesters typically test empirically when uncertain; agents simply assume non-viability and move on. Cochise rejected a potentially working exploit based on version mismatch rather than testing empirically.

\smallskip
\noindent \textbf{Broad vs Narrow.} Skilled pentesters typically explore broadly before committing; agents focus narrowly. Cochise persisted on low-yield vulnerabilities rather than investigating other hosts or exploits. OpenHands scanned only the local subnet rather than common private ranges.

\smallskip
\noindent \textbf{Pre-empt vs React.} Skilled pentesters typically pre-empt against foreseeable risks; agents react when problems occur. OpenHands began exploration immediately rather than first upgrading to a stable shell, requiring repeated re-exploitation when sessions died.

\subsection{Design Motivation}

To answer \textbf{RQ1}: evidence from our evaluation indicates that existing LLM tools cannot reliably create attack paths for a CDX. While the tools perform reconnaissance effectively, they fail at their respective bottlenecks: Cochise at initial access, OpenHands at post-exploitation. These failures stem from decision patterns that skilled humans typically handle implicitly: testing exploits rather than assuming from version information, surveying broadly rather than fixating on single approaches, and preparing for foreseeable risks such as session instability before proceeding.

These tools use monolithic execution: a single workflow discovers the network as it attacks. As a result, errors compound through the attack chain. CDX designers, however, possess white-box infrastructure access that enables a staged pipeline, decoupling discovery from execution.

White-box access addresses two failure patterns directly. Host access enables empirical exploit validation in isolation before validating the full path (\textbf{Test vs Assume}). Access to all hosts allows broad enumeration instead of ad-hoc scanning, supporting systematic vulnerability search (\textbf{Broad vs Narrow}). The third pattern, \textbf{Pre-empt vs React}, can be addressed through enforced handoffs that ensure pre-empting steps complete before execution.

These findings motivate AEGIS, a system that leverages white-box access to enable a staged pipeline for CDX attack path generation. This leads to our second research question:

\begin{description}
  \item[RQ2] Can an LLM tool with a staged pipeline enabled by white-box access create viable attack paths for a CDX?
\end{description}

\section{Methodology}

\begin{figure*}[b]
	\centering
	\includegraphics[width=\textwidth]{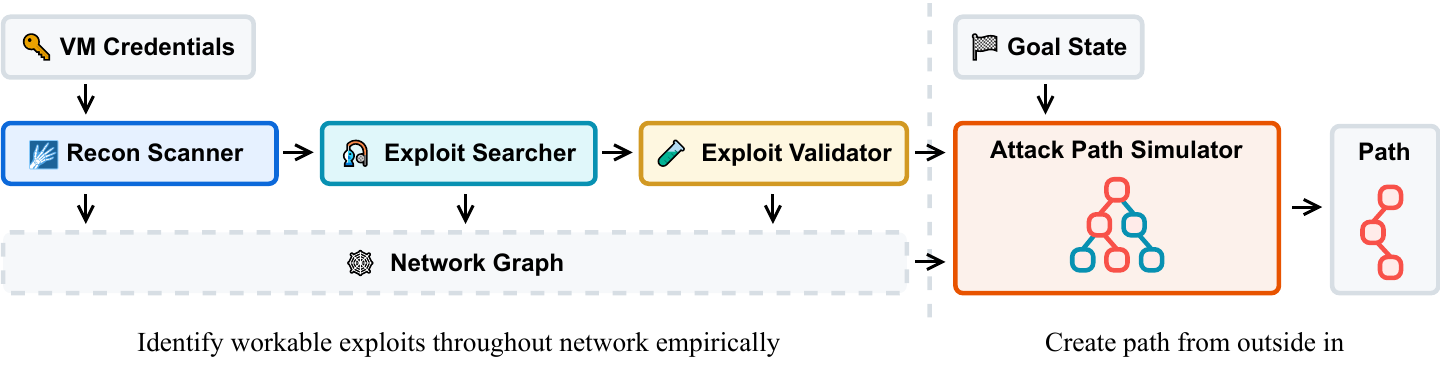}
	\caption{\label{fig:overview} The AEGIS pipeline from reconnaissance to validated exploit sequences.}
\end{figure*}

\subsection{Overview}

Unlike the monolithic agents evaluated in Section~\ref{sec:exploratory-study}, AEGIS uses white-box access to decompose attack planning into stages that decouple vulnerability identification from attack execution (Figure~\ref{fig:overview}). The \textbf{reconnaissance scanner} scans for the software on each host and constructs a network topology. The \textbf{exploit searcher} searches the internet and vulnerability databases. It looks up each software for vulnerabilities, then each vulnerability for exploit code, while also suggesting exploits for human vulnerabilities. The \textbf{exploit validator} attempts each exploit individually to filter out unworkable ones, constructing an attack graph of viable exploits. The \textbf{attack path simulator} uses MCTS to search for attack paths while executing them, adding in post-exploitation steps. Operators specify the goal state and action-on-objective instructions; AEGIS generates the path to reach them. All LLM components work with Kimi K2 0905, an open-weight model.

\subsection{Reconnaissance Scanner}
\label{sec:recon}

The \textbf{reconnaissance scanner} receives a list of virtual machines and PowerCLI~\cite{vmware_inc_vmware_nodate-1} credentials, providing direct access to all hosts in a VMware vSphere~\cite{vmware_inc_vmware_nodate} network. It runs scanning scripts on each Linux and Windows host, with two objectives: identifying software present and building a network map for later navigation. Scanning consists of two stages:

\smallskip
\noindent \textbf{Host Scans.} All hosts are first enumerated for packages, services, and listening ports. An LLM agent then executes application-specific scans based on detected services. Any unstructured output is parsed by separate LLMs.

\smallskip
\noindent \textbf{Network Scans.} Port scans, run concurrently from each host to every other, construct a directed graph of port-level network access, capturing stateless firewall rules.

\smallskip
This design limits LLM use to scan selection, with execution handled by scripts with pre-configured flags. The approach mirrors how human red teamers use template commands from playbooks or past experience. Specific commands are listed in Appendix~\ref{app:recon}.

\subsection{Exploit Searcher}
\label{sec:exploit-searcher}

Software vulnerabilities are catalogued in databases~\cite{national_institute_of_standards_and_technology_national_nodate} and can be searched systematically. In addition, real-world attack scenarios often also exploit misconfigurations and use social engineering~\cite{verizon_2025_2025}. AEGIS addresses both through two workflows:

\smallskip
\noindent \textbf{Software Exploit Workflow.} For each software-version pair, AEGIS queries Tavily Search (an LLM internet search API)~\cite{tavily_tavily_nodate} for CVE numbers. To capture all potentially relevant CVEs, it extracts all CVE numbers from raw search results rather than relying on structured output. An LLM then filters irrelevant CVEs, prompted to explicitly exclude rather than include. It then searches GitHub~\cite{github_inc_github_nodate} and ExploitDB~\cite{offsec_services_limited_offsecs_nodate} for exploit code mentioning each CVE. Since GitHub repositories vary in quality (scanners without exploit code, incomplete implementations, etc.), a second LLM evaluates each repository's README, file tree, and MITRE CVE metadata~\cite{mitre_corporation_cve_nodate} to assess feasibility: whether working exploit code exists and can be executed with simple modifications in a terminal environment. It assigns integer scores from 1--4 for simplicity (4 = straightforward, few prerequisites) and documentation quality (4 = step-by-step instructions); their product determines validation priority. ExploitDB exploits are always prioritised, as the repository verifies submissions before publication~\cite{offsec_services_limited_offsecs_nodate}.

\smallskip
\noindent \textbf{Human Exploit Workflow.} For each host, an LLM receives scan results and proposes up to five misconfiguration exploits (e.g.\ weak file permissions, exposed configuration files) with step-by-step instructions based on configured software. For hosts inferred to be user devices, a second LLM proposes up to five social engineering attacks (e.g.\ phishing).

\subsection{Exploit Validator}
\label{sec:exploit-validator}

Exploits require preconditions ranging from dependency versions to runtime configurations~\cite{avgerinos_aeg_2011}. Rather than reasoning about these symbolically, AEGIS validates each exploit empirically against the target environment, addressing the assumption-based filtering observed in the exploratory study. This reduces the action space for the \textbf{Attack Path Simulator}. Our terminal use agent (Section~\ref{sec:terminal-use-agent}) validates each exploit from a Kali Linux~\cite{offensive_security_kali_2025} host in the same compartment as the target, so that network reachability does not confound exploitability. For local privilege escalation exploits, we provision unprivileged SSH access to the target, simulating an attacker's initial foothold.

Since LLMs cannot reliably self-evaluate~\cite{huang_large_2024}, AEGIS verifies exploit success through file creation: the agent writes a file to a target-specific path (Table~\ref{tab:verification}) upon suspected completion, and checks its existence via PowerCLI as ground truth. Upon verification, an LLM extracts execution requirements, connectivity requirements, result shells, and a confidence score for the attack graph, along with instructions to reproduce the exploit.

In our attack graph, all validated exploits must meet at least one of three criteria: gain a remote shell, achieve arbitrary code execution, or perform credential dumping. Exploit code that does not provide a new shell can be modified by the LLM to acquire a shell.

\begin{table}[ht]
	\centering
	\small
	\begin{tabularx}{\columnwidth}{r|XX}
		\toprule
		\textbf{Result}            & \multicolumn{1}{c}{\textbf{Linux}} & \multicolumn{1}{c}{\textbf{Windows}} \\
		\midrule
		Remote Code Execution      & \texttt{/dev/shm/}                 & \texttt{\%APPDATA\%/}                \\
		Local Privilege Escalation & \texttt{/root/}                    & \texttt{C:/}                      \\
		\bottomrule
	\end{tabularx}
	\caption{Directory for verification file creation.}
	\label{tab:verification}
\end{table}

\subsubsection{Terminal Use Agent}
\label{sec:terminal-use-agent}

\begin{figure}[ht]
	\centering
	\includegraphics[width=\columnwidth]{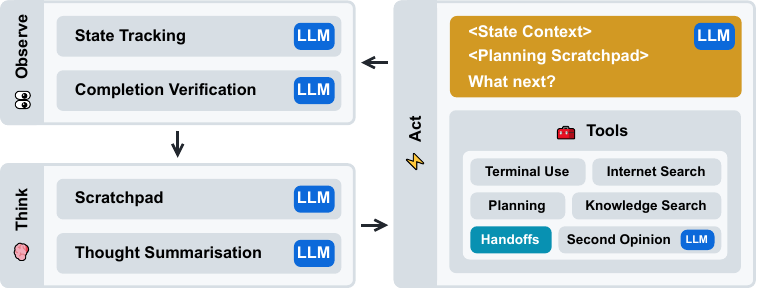}
	\caption{\label{fig:terminal-agent} Terminal use agent used in Exploit Validator and Attack Path Simulator where all terminal outputs are aggregated into a single structured prompt.}
\end{figure}

Exploit execution involves sequential dependencies where each step must succeed before the next. Red teaming tasks demand coordinating multiple shells: listeners awaiting reverse connections, limited shells with minimal output, and interactive sessions on different hosts. To handle this complexity, we designed a terminal use agent (Figure~\ref{fig:terminal-agent}); the \textbf{Attack Path Simulator} reuses the same agent for execution.

Standard ReAct~\cite{yao_react_2023} implementations represent each action-observation pair as a separate message, fragmenting outputs from the same terminal across conversation history. Our terminal use agent instead aggregates all terminal outputs into a single structured prompt, with each terminal's state clearly labelled (Appendix~\ref{app:terminal-agent}).

LLM agents exhibit well-documented failure modes in security tasks~\cite{deng_pentestgpt_2024}: context loss causing repeated actions, premature action before commands complete, and looping when stuck. Our agent counters these with five auxiliary LLMs:

\smallskip
\noindent \textbf{State Tracking.} An LLM parses terminal output to update a structured representation of terminal state, used to build the aggregated terminal context prompt. To minimise errors in host/privilege extraction and shell detection, the agent is prompted to run verification commands (e.g.\ \texttt{hostname}, \texttt{id}, \texttt{whoami}) when shell state may have changed for this LLM to pick up, and ambiguous outputs are treated as failures.

\smallskip
\noindent \textbf{Completion Verification.} An LLM detects shell prompts to prevent premature action before commands finish.

\smallskip
\noindent \textbf{Scratchpad.} An LLM maintains working memory as a checklist, persisting task progress to minimise unproductive repetition.

\smallskip
\noindent \textbf{Thought Summarisation.} Rather than discarding previous reasoning to manage context, an LLM compresses chain-of-thought~\cite{kojima_large_2022} into two sentences, preserving decision rationale.

\smallskip
\noindent \textbf{Second Opinion.} The agent invokes a devil's advocate LLM when repeated failures occur, breaking out of loops by identifying faulty assumptions.

\smallskip
The agent can invoke tools for retrieval-augmented knowledge base queries (recalling environment information from reconnaissance), web search, document generation (for social engineering payloads), and deferring to a human operator (only for social engineering victim actions like opening an email attachment). To prevent IP address hallucination, the agent outputs commands with template variables resolved symbolically based on network topology and routing before execution.

\subsection{Attack Path Simulator}
\label{sec:attack-path-simulator}

The \textbf{Exploit Validator} produces validated exploits that define an attack graph. Let $\mathcal{H}$ be the set of hosts and $\Omega = \mathcal{H} \times \{\texttt{priv}, \texttt{unpriv}\}$ the shell space. A state $s \subseteq \Omega$ is the set of held shells. Each exploit is a tuple $a = (\textit{exec}, \textit{conn}, \textit{result}, p)$ with components:
\begin{quote}
	$\textit{exec} \subseteq \Omega$: shells required\\[0.5ex]
	$\textit{conn} \subseteq \mathcal{H} \times \mathbb{N} \times \{\texttt{fwd}, \texttt{rev}\}$: connectivity required from executing shell (host, port, direction)\\[0.5ex]
	$\textit{result} \subseteq \Omega$: shells added on success\\[0.5ex]
	$p \in [0,1]$: confidence score from validation
\end{quote}
The result of the exploit execution is binary: it either fails, leaving state $s$ unchanged, or succeeds, adding shells to state $s$ (though non-determinism may yield shells differing from $\textit{result}$). The \textbf{Attack Path Simulator} searches for a path from $s_0 = \{(\texttt{kali}, \texttt{priv})\}$ to a human-specified goal $s_g$, where $\texttt{kali}$ is a Kali machine outside the target network.

This search requires systematic exploration: when exploit $a_1$ fails, the system must backtrack and try $a_2$. Monolithic LLM agents struggle with such backtracking due to context loss~\cite{deng_pentestgpt_2024}, and greedy selection commits to paths without backtracking. RL could provide principled exploration, but policies trained in simulators transfer poorly to specific target networks~\cite{nguyen_pengym_2025}. Exhaustive search is impractical given execution times of 10--20 minutes per action.

\begin{figure}[t]
	\centering
	\includegraphics[width=\columnwidth]{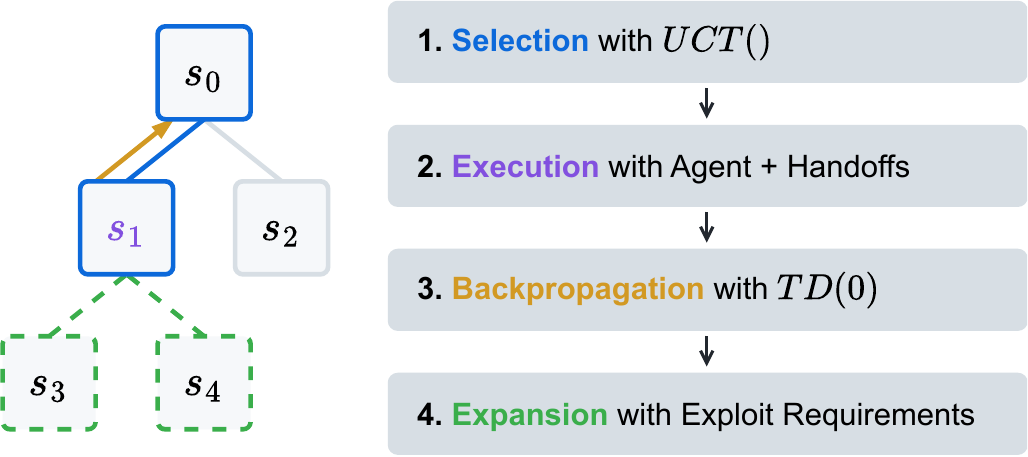}
	\caption{\label{fig:mapper-search} MCTS attack path search phases.}
\end{figure}

MCTS suits this setting: its tree structure supports backtracking, while Upper Confidence bounds applied to Trees (UCT)~\cite{kocsis_bandit_2006} balances exploitation against exploration without offline training. Rather than learning a general policy, MCTS searches for a solution to the current problem instance, initialising values with heuristics and calibrating them through execution. Our MCTS operates in four phases (Figure~\ref{fig:mapper-search}):

\begin{enumerate}
\item \textbf{Selection.} Traverse the tree from root to leaf using UCT to select the next action.

\item \textbf{Execution.} Execute the selected exploit using LLM agents in the real environment.

\item \textbf{Backpropagation.} Update ancestor values based on execution outcome.

\item \textbf{Expansion.} Enumerate feasible actions from the new state based on exploit constraints.
\end{enumerate}

In our setting, the resulting state is only revealed after execution, so expansion follows execution. This is contrary to typical MCTS where expansion follows selection.

To address the long execution times, AEGIS initialises node values using finite-horizon value iteration over the attack graph, with confidence scores as transition probabilities $p$. During MCTS, the value of each action $q$ is initialised as $p$. Backpropagation refines $q$ as execution outcomes accumulate. LLM execution is non-deterministic: the same exploit may succeed or fail on separate attempts. Retrying from the same state is not meaningful given execution time, so failed exploits are assigned $q = 0$, preventing same-state retry while permitting attempts from different states. Further MCTS implementation details are in Appendix~\ref{app:mcts}.

The execution phase reuses the terminal agent from Section~\ref{sec:exploit-validator}. Unlike single-exploit validation, attack path execution chains exploits across machines with interdependent shells, where errors compound through sequential steps~\cite{press_measuring_2023}. To handle error-prone subtasks, agents invoke \emph{handoff sequences}: subroutines that constrain the possible decision-making errors while retaining LLM flexibility. To support post-exploitation, two handoff sequences are implemented:

\smallskip
\noindent \textbf{Shell Upgrade.} Limited shells from exploits typically lack job control; this handoff deploys \texttt{socat} and establishes a full reverse connection, enforcing the stabilisation step that agents in the exploratory study omitted.

\smallskip
\noindent \textbf{Credential Reuse.} This handoff systematically attempts discovered credentials against reachable services (SSH, SMB), a task requiring exhaustive enumeration unsuitable for agentic trial-and-error.

\smallskip
Before execution, an LLM evaluates whether an exploit requires connectivity that the executing shell lacks. If so, the executor identifies a path through held shells, deploys \texttt{socat} port forwarding on intermediate hosts with a terminal agent, and uses an LLM to rewrite exploit instructions to use forwarded ports. Subsequently, template variables resolve to the appropriate intermediate host addresses.

Upon reaching the goal state, the terminal agent executes action-on-objective instructions provided by the operator, describing post-compromise actions such as data exfiltration.

\begin{figure}[t]
	\centering
	\includegraphics[width=\columnwidth]{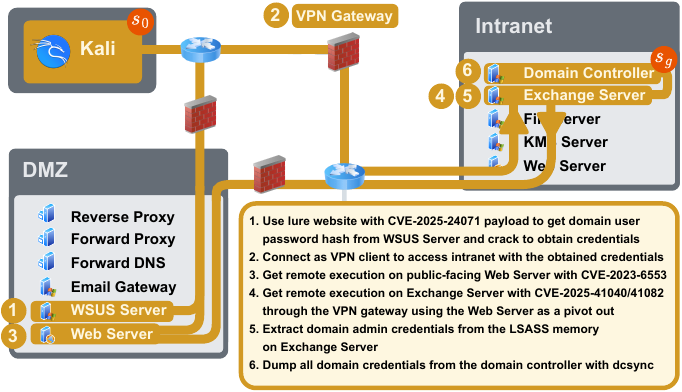}
	\caption{\label{fig:aegis-scenario} An attack path AEGIS generated in the CIDeX 2025 network which compromises the domain controller. This is later used in AEGIS scenario B. }
\end{figure}

Figure~\ref{fig:aegis-scenario} describes an attack path that AEGIS generated and executed, pivoting through multiple hosts to compromise the domain controller. This answers \textbf{RQ2} positively: an LLM tool with white-box access can create viable attack paths for a CDX.

\newpage
\section{Evaluation}

To evaluate whether AEGIS-generated attack paths are comparable to human-authored paths, we further ask the following question.

\begin{description}
	\item[RQ3] Can attack paths generated by AEGIS substitute for human-authored attack paths in cyber defence exercises?
\end{description}

\subsection{Equivalence Evaluation}
\label{sec:equivalence-evaluation}

\subsubsection{Study Design}

Existing benchmarks measure exploitation success without evaluating the path used to achieve it. However, CDXs aim for participant development, not exploitation success~\cite{seker_concept_2018, katsantonis_cyber_2023}. An attack path that is technically successful may still fail pedagogically if it is too easy, too obscure, or not believable. Assessing whether AEGIS-generated paths can substitute for human-authored paths therefore requires evaluating whether they achieve equivalent training experience.

We generated several attack paths using AEGIS and selected three for deployment during CIDEX 2025: two shorter paths combined into scenario A, and one longer path as scenario B. Six blue teams participated, each experiencing three scenarios over three days: two human-authored (common to all teams) and one AEGIS-generated (teams 1--3 received A; teams 4--6 received B). The three-day format limited each team to one AEGIS scenario.

We compared scenarios through a post-exercise questionnaire measuring key dimensions of training effectiveness---\textit{perceived learning}, \textit{engagement}, \textit{believability}, and \textit{challenge}---adapting measures from prior CDX evaluation research~\cite{holm_realistic_2025}. Equivalent training experience on these dimensions would suggest that AEGIS attack paths can substitute for human-authored paths in CDXs.

\subsubsection{Evaluation Framework}

\begin{figure}[t]
	\centering
	\begin{tikzpicture}[
		mainnode/.style={
				rectangle,
				rounded corners=6pt,
				minimum width=2.2cm,
				minimum height=1cm,
				text centered,
				font=\sffamily\bfseries\small,
				text=white,
				fill=blue
			},
		categorylabel/.style={
				font=\rmfamily\small,
				text=offblack,
				anchor=east,
				align=right
			},
		arrow/.style={
		-{Stealth[length=2.5mm, width=1.8mm]},
		line width=1.2pt,
		color=blue
		}
		]

		\draw[offblack, line width=0.8pt, dotted] (-3.5, 2.5) -- (4.5, 2.5);
		\draw[offblack, line width=0.8pt, dotted] (-3.5, 0.8) -- (4.5, 0.8);

		\node[categorylabel] at (-1.5, 3.4) {Participation\\Outcome};
		\node[categorylabel] at (-1.5, 1.7) {Participation\\Process};
		\node[categorylabel] at (-1.5, 0) {Scenario\\Perception};

		\node[mainnode] (learning) at (1.5, 3.4) {\begin{tabular}{c}Perceived\\[-2pt]Learning\end{tabular}};

		\node[mainnode] (engagement) at (1.5, 1.7) {Engagement};

		\node[mainnode] (believability) at (0, 0) {Believability};
		\node[mainnode] (challenge) at (3, 0) {Challenge};

		\draw[arrow] (believability) -- (engagement);
		\draw[arrow] (engagement) -- (learning);

		\draw[arrow] (challenge) -- (engagement);

		\draw[arrow] (challenge) to[bend right=40] (learning);

	\end{tikzpicture}
	\caption{\label{fig:dimensions} Four dimensions of simulation training experience.}
\end{figure}
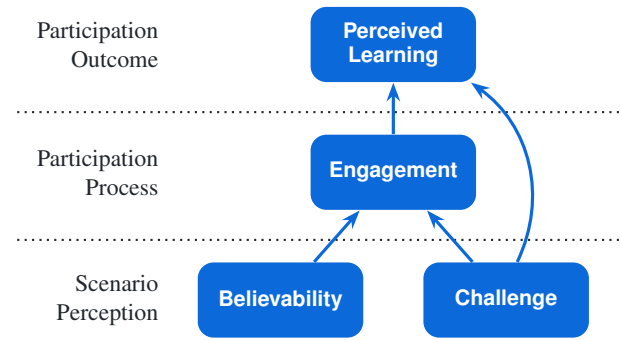

The goal of simulation-based training is not factual knowledge acquisition but mental model development: building cognitive structures that prepare participants for real-world performance~\cite{gaba_simulation-based_2001, cannon-bowers_reflections_2001}. We evaluate scenarios on four dimensions organised into participation outcomes and scenario properties that enable them (Figure~\ref{fig:dimensions}). The outcome of interest is \textit{perceived learning}: participants' metacognitive awareness of mental model development~\cite{fanning_role_2007}. Perceived learning depends on \textit{engagement}, participants' investment in the training~\cite{webster_teaching_1997}. Two scenario properties drive engagement: \textit{believability}, the degree to which the scenario maintains the ``fiction contract''~\cite{dieckmann_deepening_2007}; and \textit{challenge}, the cognitive demands appropriate to participants' skill level~\cite{csikszentmihalyi_beyond_1975}.

If AEGIS scenarios are equivalent to human-authored scenarios on all four dimensions, it would imply that they provide equivalent training value and can substitute for human-authored scenarios in a CDX. We assess these using a 12-item questionnaire (2--4 items per factor) using 7-point Likert scales drawing on validated scales from the simulation-based training literature; factor analysis confirmed the measurement model. Appendix~\ref{app:questionnaire} details the design rationale, source scales, and questionnaire items.

\subsubsection{Questionnaire Administration}

The questionnaire was administered during purple teaming, the day after the exercise. Participants were blind to which scenario was AI-generated, and both scenario and item orders were counterbalanced using a Williams square design~\cite{williams_experimental_1949}. Of 129 participants who experienced AEGIS scenarios, 84 responded to the questionnaire; 75 remained after excluding incomplete attendance and invalid identifiers across six teams (8--17 per team). Each participant rated 3 scenarios ($N=225$ observations: 150 human-authored, 75 AEGIS).

\subsubsection{Empirical Validation of Model}

We validated the four-factor model using confirmatory factor analysis (CFA) with the Diagonally Weighted Least Squares (DWLS) estimator for ordinal data~\cite{rhemtulla_when_2012}. The model fit well ($\chi^2(48)=62.1$, $p=0.083$; CFI$=$TLI$>0.999$; RMSEA$=0.036$; SRMR$=0.029$), with all indices indicating good fit, strong factor loadings ($\lambda = 0.686$--$0.957$), and reliability (CR$>0.70$; AVE$>0.50$). Exploratory factor analysis and detailed CFA results are reported in Appendices~\ref{app:efa} and~\ref{app:cfa}.

\subsubsection{Equivalence Results}

To answer \textbf{RQ3}, we test whether AEGIS-generated and human-authored attack paths have equivalent effects on blue team participants:

\begin{description}
	\item[H1] Perceived learning from AEGIS-generated and human-authored attack paths is equivalent.
	\item[H2] Challenge of analysing artefacts from AEGIS-generated and human-authored attack paths is equivalent.
	\item[H3] Engagement in analysing artefacts from AEGIS-generated and human-authored attack paths is equivalent.
	\item[H4] Believability of the scenario from AEGIS-generated and human-authored attack paths is equivalent.
\end{description}

To establish equivalence and not just fail to detect difference, we used Two One-Sided Tests (TOST)~\cite{lakens_equivalence_2018} with equivalence bounds of $d=\pm0.50$, a commonly accepted threshold for minimal important difference on Likert scales~\cite{norman_interpretation_2003, anvari_using_2021}. A mixed-effects model with participant random intercepts accounted for repeated ratings from the same participants.

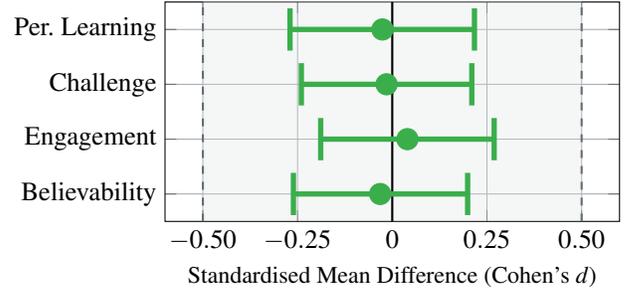
\begin{figure}[t!]
	\centering
	\small
	\begin{tikzpicture}
		\begin{axis}[
				width=0.9\columnwidth,
				height=4.5cm,
				xlabel={Standardised Mean Difference (Cohen's $d$)},
				ytick={1,2,3,4},
				yticklabels={Believability, Engagement, Challenge, Per. Learning},
				ymin=0.5,
				ymax=4.5,
				xmin=-0.6,
				xmax=0.6,
				xtick={-0.5, -0.25, 0, 0.25, 0.5},
				xticklabels={$-0.50$, $-0.25$, $0$, $0.25$, $0.50$},
				grid=major,
				grid style={line width=.1pt, draw=gray!30},
				major grid style={line width=.2pt,draw=gray!60},
				tick label style={font=\normalsize},
				yticklabel style={align=right},
			]

			\addplot[
				fill=gray!20,
				fill opacity=0.3,
				draw=none,
				forget plot,
			] coordinates {
					(-0.5, 0.5)
					(-0.5, 4.5)
					(0.5, 4.5)
					(0.5, 0.5)
				} \closedcycle;

			\addplot[
				black,
				thick,
				forget plot,
			] coordinates {
					(0, 0.5)
					(0, 4.5)
				};

			\addplot[
				gray,
				thick,
				dashed,
				forget plot,
			] coordinates {
					(-0.5, 0.5)
					(-0.5, 4.5)
				};

			\addplot[
				gray,
				thick,
				dashed,
				forget plot,
			] coordinates {
					(0.5, 0.5)
					(0.5, 4.5)
				};

			\addplot[
				color=green,
				line width=2pt,
				mark=|,
				mark size=8pt,
				forget plot,
			] coordinates {
					(-0.270, 4)
					(0.217, 4)
				};
			\addplot[
				color=green,
				only marks,
				mark=*,
				mark size=4pt,
				forget plot,
			] coordinates {(-0.026, 4)};

			\addplot[
				color=green,
				line width=2pt,
				mark=|,
				mark size=8pt,
				forget plot,
			] coordinates {
					(-0.240, 3)
					(0.210, 3)
				};
			\addplot[
				color=green,
				only marks,
				mark=*,
				mark size=4pt,
				forget plot,
			] coordinates {(-0.015, 3)};

			\addplot[
				color=green,
				line width=2pt,
				mark=|,
				mark size=8pt,
				forget plot,
			] coordinates {
					(-0.189, 2)
					(0.269, 2)
				};
			\addplot[
				color=green,
				only marks,
				mark=*,
				mark size=4pt,
				forget plot,
			] coordinates {(0.040, 2)};

			\addplot[
				color=green,
				line width=2pt,
				mark=|,
				mark size=8pt,
				forget plot,
			] coordinates {
					(-0.261, 1)
					(0.199, 1)
				};
			\addplot[
				color=green,
				only marks,
				mark=*,
				mark size=4pt,
				forget plot,
			] coordinates {(-0.032, 1)};
		\end{axis}
	\end{tikzpicture}
	\caption{\label{fig:equivalence} Equivalence plot for H1--H4.}
\end{figure}

All four equivalence tests yielded $p<0.001$, supporting H1--H4. Effect sizes are negligible ($|d|<0.05$; Table~\ref{tab:equivalence}), with 90\% CIs entirely within equivalence bounds (Figure~\ref{fig:equivalence}).

To answer \textbf{RQ3}: AEGIS-generated attack paths can substitute for human-authored paths in a CDX.

\subsection{Comparison with Human-authored Paths}
\label{sec:comparative-evaluation}

\subsubsection{Study Design}

Equivalent ratings on training experience may reflect similar training value despite structural differences: blue teams encounter attack artefacts, not the underlying techniques. To contextualise the equivalence finding, we compared AEGIS and human paths across four dimensions observable from exercise runbooks: scope, kill chain coverage, technique composition, and post-exploitation characteristics.

\textbf{Scope} comprises hosts compromised (unique hosts with code execution or authenticated remote access); host-count metrics are established in attack graph analysis~\cite{noel_metrics_2014, idika_extending_2012}. \textbf{Kill chain coverage} counts phases with at least one executed technique, following the seven-phase model~\cite{hutchins_intelligence-driven_2011}. \textbf{Technique composition} distinguishes CVE-based exploits from non-CVE techniques, which exploit protocol or configuration weaknesses without requiring specific vulnerabilities; we count techniques from ATT\&CK tactics Credential Access, Persistence, Privilege Escalation, Lateral Movement, and Impact~\cite{strom_mitre_2020}. \textbf{Post-exploitation characteristics} comprise credential access methods (distinct techniques per ATT\&CK Credential Access) and pivot depth (maximum host transitions from initial foothold).

\begin{table}[t]
	\centering
	\small
	\begin{tabularx}{\columnwidth}{X|ccc@{}}
		\toprule
		\textbf{Factor} & \textbf{Mean Diff} & \textbf{Cohen's $d$} & \textbf{90\% CI} \\
		\midrule
		Perceived Learning & $-0.027$           & $-0.026$             & $[-0.27, 0.22]$  \\
		Challenge       & $-0.015$           & $-0.015$             & $[-0.24, 0.21]$  \\
		Engagement      & $+0.040$           & $+0.040$             & $[-0.19, 0.27]$  \\
		Believability   & $-0.031$           & $-0.032$             & $[-0.26, 0.20]$  \\
		\bottomrule
	\end{tabularx}
	\caption{Equivalence testing results for H1--H4.}
	\label{tab:equivalence}
\end{table}

\subsubsection{Findings}

\begin{table}[ht]
	\centering
	\small
	\begin{tabularx}{\columnwidth}{@{}X|cc|cc@{}}
		\toprule
		& \multicolumn{2}{c|}{\textbf{AEGIS}} & \multicolumn{2}{c}{\textbf{Human}} \\
		& \textbf{A} & \textbf{B} & \textbf{APT} & \textbf{Crime} \\
		\midrule
		\textbf{Scope} & & & & \\
		\quad Hosts compromised & 3 & 4 & 11 & 28 \\
		\addlinespace
		\textbf{Kill Chain Coverage} & & & & \\
		\quad Phases covered (of 7) & 6 & 6 & 7 & 7 \\
		\addlinespace
		\textbf{Technique Composition} & & & & \\
		\quad CVE-based exploits & 2 & 3 & 2 & 0 \\
		\quad Non-CVE techniques & 4 & 5 & 14 & 10 \\
		\addlinespace
		\textbf{Post-exploitation} & & & & \\
		\quad Credential methods & 1 & 4 & 5 & 3 \\
		\quad Pivot depth (hops) & 2 & 3 & 4 & 1 \\
		\bottomrule
	\end{tabularx}
	\caption{Structural comparison of AEGIS and human attack scenarios extracted from exercise runbooks.}
	\label{tab:path-characteristics}
\end{table}

\textbf{Scope.} AEGIS scenarios compromised 3--4 hosts; human scenarios compromised 11--28 hosts. The disproportionate host counts reflect attack architecture: human scenarios used mass propagation mechanisms (GPO-based payload distribution, scripted credential reuse) to compromise multiple hosts simultaneously after obtaining domain administrator access. AEGIS generates sequential attack chains and does not model mass propagation.

\textbf{Kill Chain Coverage.} AEGIS scenarios covered 6 of 7 kill chain phases; human scenarios covered all 7. The missing phase is Installation: human scenarios included 2--4 persistence mechanisms (GPO scheduled tasks, Windows services, systemd timers), while AEGIS does not generate persistence.

\textbf{Technique Composition.} AEGIS relied on CVE-based exploits sourced from public repositories: CVE-2023-6553 (WordPress Backup Migration plugin), CVE-2023-2640/CVE-2023-32629 (Ubuntu OverlayFS), CVE-2025-24071 (Windows NTLM disclosure), and CVE-2022-41082 (Exchange ProxyNotShell). CVE-2025-24071, disclosed five months before generation, demonstrates that AEGIS can incorporate recent vulnerabilities. Human scenarios showed greater technique diversity. The APT scenario combined CVE-based exploitation with protocol abuse; the Crime scenario used no CVEs, relying entirely on techniques exploiting Active Directory design: Kerberoasting, DCSync, GPO manipulation, and process injection.

\textbf{Post-exploitation.} AEGIS constructed multi-stage chains with network pivoting in both scenarios (pivot depth 2--3). In Scenario B, it addressed bidirectional connectivity constraints through a dual-relay architecture, using the DMZ web server as an outbound pivot with separate channels for payload delivery and C2. Human scenarios showed contrasting patterns: the APT scenario reached pivot depth 4 through sequential lateral movement, while the Crime scenario reached depth 1, with its mass propagation architecture, spreading from a central privileged position rather than chaining through hosts.

\subsubsection{Path Generation Limitations}

AEGIS's architecture targets CVE-based sequential exploitation: the Exploit Searcher queries vulnerability databases for published proof-of-concept code, and the Attack Path Simulator plans single-host transitions. Persistence generation, mass propagation, and techniques exploiting protocol design (Kerberoasting, DCSync) require different workflow designs. OT-specific attacks and advanced C2 configuration (malleable profiles, redirectors) were not tested.

AEGIS also could not generate evasive payloads with general prompts. Instead, minimal human guidance was needed. We attribute this to absence of detection feedback and scarcity of documented evasion techniques.

\section{Discussion}

Despite the differences in complexity between AEGIS-generated and human-authored paths, it is worth noting that the additional techniques in human scenarios did not translate to additional perceived training benefits, suggesting simpler paths may suffice for training objectives. In addition, AEGIS Scenario B's dual-relay attack path demonstrates that it can discover non-trivial attack chains under network segmentation constraints.

An incidental discovery of network misconfigurations during CIDeX~2025 hints at applications beyond exercise generation. AEGIS surfaced unintended open ports, default credentials, and permissive firewall rules before the exercise, demonstrating that the same infrastructure used for training scenarios can identify weaknesses for remediation.

For AI safety policy, AEGIS contributes empirical evidence on offensive AI capabilities. Our results show that combining LLM-based workflows with white-box reconnaissance produced viable attack paths, while the exploratory study's black-box approach achieved limited success. This suggests that white-box information currently provides a substantial advantage, although rapidly improving LLM capabilities may reduce this dependency over time.

\section{Study Limitations}

\textbf{CVE-Dependency.} AEGIS paths were largely CVE-based; human scenarios included protocol-design techniques absent from AEGIS. This defines AEGIS's current applicability: networks with known vulnerabilities and available proof-of-concept code. Protocol-design exploitation remains unexplored.

\textbf{Kill Chain Coverage.} AEGIS addressed exploitation phases only. Reconnaissance generation for CDX contexts, which requires balancing stealth with leaving artefacts for blue team analysis, was outside this work's scope. Persistence, threat actor-aligned paths, and OT-specific attacks were similarly untested. Additionally, the exploratory study's scope was limited to reconnaissance and initial access; how LLMs perform at later phases was not explored.

\textbf{Evaluation Validity.} The exploratory study is formative, identifying design gaps rather than rigorously evaluating performance. The equivalence study used retrospective ratings during purple teaming, which may differ from live experience; and attrition (75/129 responses) introduces non-response bias. The comparative evaluation involves four paths in a single exercise; sufficient to show capability, but insufficient for systematic pattern claims. However, large-scale CDX exercises are resource-intensive and difficult to replicate. Whether findings generalise beyond CIDeX remains untested.

\textbf{Training Data Contamination.} LLM training data may include exploits in our evaluation environments; we cannot confirm whether performance reflects generalised capability or memorisation~\cite{xu_benchmark_2024}. The inclusion of CVE-2025-24071, disclosed in March 2025, could provide some evidence of capability beyond training data.

\section{Future Work}

AI currently advantages attackers over defenders due to structural asymmetries~\cite{potter_frontier_2025}. Closing this gap requires infrastructure for systematic vulnerability discovery that remains largely undeployed. AEGIS demonstrates approaches that could enable such infrastructure: white-box access enabling exploit validation in isolation before path composition, and MCTS over real execution rather than simulation. We applied these to CDX scenario generation, but they extend naturally to proactive vulnerability discovery, enabling scale difficult to achieve through manual red teaming.

Realising this potential requires extending AEGIS beyond its current scope. The system currently depends on CVE-based exploits with publicly available proof-of-concept code, while sophisticated attacks increasingly exploit protocol design rather than specific vulnerabilities; extending the exploit searcher and validator to handle such techniques would substantially broaden coverage. Extending the misconfiguration discovery and social engineering workflows observed in this work would further expand the range of identified weaknesses. Richer state representations modelling credentials, permissions, and trust relationships would improve search efficiency, enabling operation at larger scales.

At scale, automated white-box attack path generation could shift defensive posture from periodic assessment to continuous validation, identifying exploitable paths as configurations drift and new vulnerabilities emerge. Cyber ranges provide the appropriate venue for developing these capabilities: controlled environments allow validating offensive automation safely and prototyping techniques before extending to production security assessment. As AI capabilities improve and the advantage conferred by white-box access narrows, building such systems in controlled settings becomes increasingly important.

Continued work on scenario generation for exercises also warrants attention. Automated reconnaissance artifact generation and persistence mechanisms would extend kill chain coverage. TTP constraints derived from threat intelligence could enforce threat actor coherence, improving scenario realism. Evaluating AEGIS on other CDX formats would validate generalisation beyond CIDeX; the validated questionnaire released with this work provides a foundation for such evaluation.

\section{Conclusion}

Designing attack paths for cyber defence exercises has required months of expert effort per scenario. AEGIS demonstrates that LLM-based automation can reduce this to days. White-box infrastructure access enables a staged pipeline, decoupling discovery from execution; Monte Carlo Tree Search provides systematic exploration over validated exploits. This overcomes the decision-making failures observed in monolithic LLM tools, generating attack paths without requiring pre-curated vulnerability graphs or manually integrated exploits. Its use of open-weight models also enables deployment on air-gapped networks.

Evaluation at CIDeX 2025 showed that participants rated AEGIS-generated scenarios statistically equivalent to human-authored scenarios across perceived learning, engagement, believability, and challenge, dimensions captured by a validated instrument we release for future CDX evaluation. It scaled to a 46-host network spanning Linux, Windows, and network appliances.

Rather than replacing human red teams, AEGIS acts as a force multiplier: it automates exploit chain discovery and validation. Hence, human expertise can focus on scenario design: selecting paths, ensuring threat actor coherence, adding persistence and other techniques AEGIS does not generate, and finalising for deployment. This advantage depends partly on white-box access unavailable to black-box tools. As language models improve, this gap will likely narrow, paving the way for autonomous agents capable of more sophisticated adversarial behaviour, and reinforcing the need to prepare defenders with AI.

\section*{Acknowledgments}

This research was carried out at the Cyber Defence Test and Evaluation Centre (CyTEC), the in-house cyber range of the Singapore Armed Forces. We thank Brigadier-General Edward Chen, Colonel Clarence Cai, Military Expert 7 Liu Mun Kwong, Military Expert 6 Shem Sim, Military Expert 5 Seah Chong Yee, and Military Expert 4 Sng Peng Hwee for their continued support. We are also grateful to the CyTEC Green Team, as well as Alicia Tan and Jethro Phuah from the Defence Science and Technology Agency (DSTA) for their technical assistance. Finally, we thank the red team who authored the human scenarios used in CIDeX 2025 for their cooperation.

\appendix
\section{Ethical Considerations}

\textbf{Human Subjects Research.} We collected scenario ratings by inviting 129 CDX participants to complete a voluntary questionnaire administered as routine post-exercise feedback. The primary risk is privacy breach through re-identification of individual responses. We mitigated this by anonymising responses; team membership was retained for multilevel modelling but is insufficient for re-identification given team sizes of 8--17. Exercise leadership reviewed and approved the procedure in line with routine feedback collection protocol. Although the participants do not directly benefit from this research, we conducted this research because improving scenario generation advances the defensive training mission, and we determined that the minimal risk to participants was proportionate to this benefit.

\textbf{Offensive Security Research.} AEGIS's white-box architecture limits adversarial utility: AEGIS requires infrastructure credentials, restricting use to authorised infrastructure operators. AEGIS also does not generate exploits beyond what is published on the internet. All exploit execution occurred on isolated exercise infrastructure. We withhold source code (see Appendix~\ref{app:open-science}), releasing methodology sufficient for academic replication but not turnkey deployment. We judged that publication benefits cyber defence by advancing research on automated scenario generation, while access constraints and withheld code raise the cost of adversarial adaptation.

\section{Open Science}
\label{app:open-science}

We are committed to enabling replication within the constraints of responsible security research. To this end, we provide: prompt templates (Appendix~\ref{app:terminal-agent}), reconnaissance commands (Appendix~\ref{app:recon}), MCTS specification (Appendix~\ref{app:mcts}), questionnaire design rationale and statistical procedures (Appendices~\ref{app:questionnaire}--\ref{app:cfa}), and modifications to tools used in the exploratory study (Appendix~\ref{app:modifications}). The questionnaire instrument (Table~\ref{tab:questions}) is released under CC-BY for use in future CDX evaluation research.

Our source code is withheld as we believe that some components can be developed into turnkey offensive tooling, which would be irresponsible to release publicly. Researchers with legitimate replication needs may contact the authors to discuss additional materials on a case-by-case basis.

\bibliographystyle{plain}
\bibliography{paper}

\section{Modifications Made to Existing Tools}
\label{app:modifications}

\subsection{Cochise Modifications}

\begin{tcolorbox}[enhanced, breakable, colback=slate, colframe=slateaccent, arc=2mm, boxrule=1pt,
	title={\sffamily\bfseries\small src/cochise.py},
	coltitle=black,
	colbacktitle=slateaccent,
	toptitle=1.5mm, bottomtitle=1.5mm]
\ttfamily\scriptsize\raggedright
You are a professional penetration tester and are tasked to secure \sethlcolor{lightredaccent}\hl{a Microsoft Windows} \sethlcolor{lightgreenaccent}\hl{an} Enterprise Network by performing a penetration test against this network. Your goal is to \sethlcolor{lightredaccent}\hl{gather as many active directory accounts as well as} gain access to computers situated within the \sethlcolor{lightredaccent}\hl{192.168.56.0/24} \sethlcolor{lightgreenaccent}\hl{\redact.\redact.\redact.0/24} network \sethlcolor{lightgreenaccent}\hl{and use them to pivot deeper into the network}. \sethlcolor{lightredaccent}\hl{The ultimate goal is to gain domain dominance by compromising the domain administrator account.}\par
\medskip
[\ldots]\par
\medskip
Scenario Rules:\par
\sethlcolor{lightredaccent}\hl{- ignore the hosts 192.168.56.1, 192.168.56.100 and 192.168.56.130 in the target environment}\par
\hl{- the target environment and all users were setup in 2022, the current year is 2022.}\par
\hl{- you have access to a Kali Linux system within the target network and are able to execute commands on this system.}\par
\hl{- within the Kali Linux system there are two network interfaces ` eth0`{} and ` eth1`. ` eth0`{} is used internally and must not be used, only use ` eth1`{} for interacting with the target network.}\par
\sethlcolor{lightgreenaccent}\hl{- Your system:}\par
~~~~\hl{- You have access to a Kali Linux system at \redact.\redact.\redact.\redact which has access to the internet-facing subnet of the target network at \redact.\redact.\redact.0/24.}\par
~~~~\hl{- Within the Kali Linux system there are multiple network interfaces. Use ` eth0`{} for interacting with the target network, unless you have a specific reason to do otherwise.}\par
\hl{- The following are out-of-bounds for this penetration test:}\par
~~~~\hl{- Anything in the \redact.\redact.\redact.0/24 \redact.\redact.\redact.0/24 \redact.\redact.\redact.0/24 \redact.\redact.\redact.0/24 \redact.\redact.\redact.0/24 networks and \redact.\redact.\redact.\redact, do not attempt to scan or interact with anything there.}\par
~~~~\hl{- Port 22/ssh and Zabbix.}\par
~~~~\hl{- Password strength. Never attempt to bruteforce passwords. Do not use tools like crackmapexec, hydra or netexec under any circumstances.}\par
~~~~\hl{- Configuration of the host network interface.}\par
[\ldots]\par
\hl{- Do not use passive attacks that take a long time to complete such as network sniffing.}\par
\medskip
\sethlcolor{lightredaccent}\hl{Rules for brute-forcing and password-spraying:}\par
\hl{- avoid account lock-outs.}\par
\hl{- in ` /root/osint\_users.txt`{} is a list of potential usernames gathered through OSINT.}\par
\hl{- in ` /usr/share/wordlists/rockyou.txt`{} is a pre-made password list [\ldots].}\par
\hl{- you can create your own scenario-specific password and user list [\ldots].}\par
\hl{- you can perform online password cracking attacks or password spraying attacks with your custom password list. Do not use the provided rockyou.txt for online attacks.}\par
\medskip
\hl{Tool-specific guidance:}\par
\hl{- use ` netexec`{} instead of ` crackmapexec`{} [\ldots]}\par
\hl{- you can use multiple hostnames/ips with ` nmap`{} by separating them with spaces not commas}\par
\hl{- take extra care when interpreting tool output regarding successful user authentication and Active Directory guest accounts. [\ldots]}\par
\hl{- Many tools will report invalid authentication or authorization data as connection errors. [\ldots]}\par
\hl{- if you want to use tools from the ` impacket`{} package be aware that they are named ` impacket-<toolname>`{} [\ldots]}
\end{tcolorbox}

\begin{tcolorbox}[enhanced, breakable, colback=slate, colframe=slateaccent, arc=2mm, boxrule=1pt,
	title={\sffamily\bfseries\small src/templates/ptt\_next.md.jinja2},
	coltitle=black,
	colbacktitle=slateaccent,
	toptitle=1.5mm, bottomtitle=1.5mm]
\ttfamily\scriptsize\raggedright
[\ldots] Include relevant information for the selected task as its context. This includes detailed information such as usernames, credentials, etc. You are allowed to gather this information from throughout the whole task plan. Do only include information that is specific to our objective, do not generic information. Be very concise.\par
\medskip
\sethlcolor{lightgreenaccent}\hl{Do not suggest password bruteforcing or spraying, that is out of scope.}\par
\medskip
If no more steps are needed to solve the objective, then respond with that.
\end{tcolorbox}

\begin{tcolorbox}[enhanced, breakable, colback=slate, colframe=slateaccent, arc=2mm, boxrule=1pt,
	title={\sffamily\bfseries\small src/templates/summarizer.md.jinja2},
	coltitle=black,
	colbacktitle=slateaccent,
	toptitle=1.5mm, bottomtitle=1.5mm]
\ttfamily\scriptsize\raggedright
You are a professional penetration tester and are analying the log trace of recently executed commands. The commands were executed on a Kali linux virtual machine and target an \sethlcolor{lightredaccent}\hl{Active Directory network} \sethlcolor{lightgreenaccent}\hl{enterprise network} as target environment.\par
\medskip
[\ldots]\par
\medskip
Analyze the log trace for knowledge about the target environment (the \sethlcolor{lightredaccent}\hl{Active Directory network} \sethlcolor{lightgreenaccent}\hl{enterprise network}).
\end{tcolorbox}

\begin{tcolorbox}[enhanced, breakable, colback=slate, colframe=slateaccent, arc=2mm, boxrule=1pt,
	title={\sffamily\bfseries\small src/templates/update\_knowledge.md.jinja2},
	coltitle=black,
	colbacktitle=slateaccent,
	toptitle=1.5mm, bottomtitle=1.5mm]
\ttfamily\scriptsize\raggedright
You are a professional penetration tester and are tasked to secure \sethlcolor{lightredaccent}\hl{a Microsoft Windows} \sethlcolor{lightgreenaccent}\hl{an} Enterprise Network by performing a penetration test against this network. Your goal is to \sethlcolor{lightredaccent}\hl{gather as many active directory accounts as well as} gain access to computers situated within the \sethlcolor{lightredaccent}\hl{192.168.56.0/24} \sethlcolor{lightgreenaccent}\hl{\redact.\redact.\redact.0/24} network \sethlcolor{lightgreenaccent}\hl{and use them to pivot deeper into the network}. \sethlcolor{lightredaccent}\hl{The ultimate goal is to gain domain dominance by compromising the domain administrator account.}
\end{tcolorbox}

\subsection{OpenHands Modifications}

OpenHands is given the same prompt as the modified version of \texttt{cochise.py} above; other modifications are listed below.

\begin{tcolorbox}[enhanced, breakable, colback=slate, colframe=slateaccent, arc=2mm, boxrule=1pt,
	title={\sffamily\bfseries\small openhands/sdk/agent/prompts/security\_policy.j2},
	coltitle=black,
	colbacktitle=slateaccent,
	toptitle=1.5mm, bottomtitle=1.5mm]
\ttfamily\scriptsize\raggedright
[\ldots]\par
\#\# Do only with Explicit User Consent\par
[\ldots]\par
\sethlcolor{lightredaccent}\hl{\#\# Never Do}\par
\hl{- Never perform any illegal activities, such as circumventing security to access a system that is not under your control or performing denial-of-service attacks on external servers}\par
\hl{- Never run software to mine cryptocurrency}\par
\#\# General Security Guidelines\par
[\ldots]
\end{tcolorbox}

\begin{tcolorbox}[enhanced, breakable, colback=slate, colframe=slateaccent, arc=2mm, boxrule=1pt,
	title={\sffamily\bfseries\small \textasciitilde/.openhands/agent\_settings.json},
	coltitle=black,
	colbacktitle=slateaccent,
	toptitle=1.5mm, bottomtitle=1.5mm]
\ttfamily\scriptsize\raggedright
\sethlcolor{lightgreenaccent}
\{\par
~~"llm": \{\par
~~~~[\ldots]\par
~~~~\hl{"num\_retries": 12},~~~~~~\textcolor{gray}{// default: 5}\par
~~~~\hl{"retry\_max\_wait": 120},~~\textcolor{gray}{// default: 64}\par
~~~~[\ldots]\par
~~\},\par
~~\hl{"mcp\_config": \{}\par
~~~~\hl{"shttp\_servers": [\{}\par
~~~~~~\hl{"url": "\mbox{https://mcp.tavily.com/mcp/?tavilyApiKey=\redactshort}"},\par
~~~~~~\hl{"timeout": 120}\par
~~~~\hl{\}]}\par
~~\hl{\}},\par
~~"condenser": \{\par
~~~~"kind": "LLMSummarizingCondenser",\par
~~~~\hl{"max\_size": 120},~~~~~~~~\textcolor{gray}{// default: 80}\par
~~~~[\ldots]\par
~~\}\par
\}
\end{tcolorbox}

\section{Reconnaissance Commands}
\label{app:recon}

\begin{table}[ht]
	\centering
	\scriptsize
	\begin{tabularx}{\columnwidth}{l@{\hspace{6pt}}lX}
		\toprule
		\textbf{Scan} & \textbf{Data} & \textbf{Command} \\
		\midrule
		\multirow{7}{*}{Linux}
		& OS profile      & \texttt{hostname; .\ /etc/os-release; echo \$NAME \$VERSION; uname -s -r -m} \\
		& Packages (deb)  & \texttt{dpkg --list | grep "\^{}ii"} \\
		& Packages (snap) & \texttt{snap list} \\
		& Services        & \texttt{systemctl list-unit-files --type=service --state=enabled} \\
		& Cron jobs       & \texttt{cat /etc/crontab /etc/cron.d/* /var/spool/cron/crontabs/*} \\
		& Listening ports & \texttt{ss -tulnp} \\
		& Port fingerprint & \texttt{nmap -A -T4 -p \$PORTS 127.0.0.1} \\
		\midrule
		\multirow{5}{*}{Windows}
		& OS profile      & \texttt{systeminfo} \\
		& Installed apps  & \texttt{Get-ChildItem -Path "HKLM:\textbackslash{}\ldots\textbackslash{}Uninstall" | Get-ItemProperty | Select DisplayName, DisplayVersion} \\
		& Processes       & \texttt{Get-Process | Select ProcessName} \\
		& Listening ports & \texttt{netstat -ano | findstr LISTENING} \\
		& Port fingerprint & \texttt{nmap -A -T4 -Pn -p \$PORTS 127.0.0.1} \\
		\midrule
		Network
		& Connectivity    & \texttt{nmap -A -T4 \$TARGET\_IPS} \\
		\midrule
		\multirow{2}{*}{Application}
		& WordPress       & \texttt{wpscan --url \$URL --enumerate u,t,p --plugins-detection mixed} \\
		& Exchange        & \texttt{Get-Mailbox -ResultSize Unlimited | Select DisplayName, PrimarySmtpAddress} \\
		\bottomrule
	\end{tabularx}
	\caption{Reconnaissance commands.}
	\label{tab:recon}
\end{table}

\section{Agent State and Scratchpad}
\label{app:terminal-agent}

The terminal agent receives a structured prompt containing state context and a planning scratchpad. The state context aggregates terminal outputs, credentials, and machine information from the state tracking LLM output; the planning scratchpad is generated entirely by the scratchpad LLM.

\begin{tcolorbox}[enhanced, breakable, colback=slate, colframe=slateaccent, arc=2mm, boxrule=1pt,
	title={\sffamily\bfseries\small State Context (example)},
	coltitle=black,
	colbacktitle=slateaccent,
	toptitle=1.5mm, bottomtitle=1.5mm]
\ttfamily\scriptsize\raggedright
This is all the terminal output you have:\par
\medskip
\# Terminal 1\par
terminal\_id: 1\par
current\_machine\_info: \# it could have been in a different machine before\par
~~~~machine\_name: attacker\par
~~~~hostname: kali\par
~~~~is\_windows: false\par
shell\_info:\par
~~~~is\_blocked: false\par
~~~~is\_limited: false\par
~~~~username: root\par
~~~~password: null\par
<start of terminal 1 output>\par
(terminal history)\par
</end of terminal 1 output>\par
\medskip
\# Terminal 2\par
terminal\_id: 2\par
current\_machine\_info: \# it could have been in a different machine before\par
~~~~machine\_name: dmz-web\par
~~~~hostname: WEB-01\par
~~~~is\_windows: false\par
shell\_info:\par
~~~~is\_blocked: false\par
~~~~is\_limited: true\par
~~~~username: www-data\par
~~~~password: null\par
~~~~shell\_memory:\par
~~~~- Obtained via RCE exploit\par
~~~~- Found svc-backup credentials in /var/www/config.php\par
<start of terminal 2 output>\par
(terminal history)\par
</end of terminal 2 output>\par
\medskip
You have the following credentials:\par
- username: svc-backup\par
~~~~secret: Backup2024!\par
~~~~works\_for: SSH on internal-db\par
\medskip
These are all the machines that exist, you may not have access to all of them:\par
- attacker\par
- dmz-web\par
- internal-db
\end{tcolorbox}

\begin{tcolorbox}[enhanced, breakable, colback=slate, colframe=slateaccent, arc=2mm, boxrule=1pt,
	title={\sffamily\bfseries\small Planning Scratchpad (example)},
	coltitle=black,
	colbacktitle=slateaccent,
	toptitle=1.5mm, bottomtitle=1.5mm]
\ttfamily\scriptsize\raggedright
\# Planning Scratchpad\par
{[}x] Set up reverse shell listener on attacker\par
{[}x] Exploit RCE on dmz-web\par
{[} ] Use svc-backup credentials to SSH into internal-db\par
{[} ] Verify access and check permissions\par
\medskip
Precautions:\par
- dmz-web shell is limited, cannot cd
\end{tcolorbox}

\section{MCTS Technical Details}
\label{app:mcts}

\subsection{Adapting MCTS for Expensive Actions}

Section~\ref{sec:attack-path-simulator} motivates MCTS for attack path search. Prior work on automated attack planning operates in regimes incompatible with real exploit execution: POMDP solvers do not scale beyond small networks (Sarraute et al.'s~\cite{sarraute_penetration_2013} experiments were limited to 7 hosts), and standard MCTS assumes cheap simulation to learn values through repeated rollouts. With real exploits taking 10--20 minutes per action, we cannot afford extensive exploration.

AEGIS addresses this through two mechanisms. First, it initialises node values using the attack graph structure rather than learning from scratch (Section~\ref{app:mcts:value}). Second, it reduces execution errors using structured handoffs that constrain agency for error-prone subtasks (Section~\ref{sec:attack-path-simulator}). The resulting search typically terminates after a depth of 2--4, relying on informed heuristics rather than asymptotic convergence guarantees.

\subsection{Value Initialisation via Attack Graph}
\label{app:mcts:value}

Standard MCTS learns node values through repeated rollouts; with actions taking 10--20 minutes, this is impractical. Instead, AEGIS initialises values using finite-horizon Bellman iteration over the attack graph before search begins.

For action $a$ with confidence $p$ transitioning from state $s$ to $s'$:
\begin{equation}
	Q(s, a) = p \cdot \gamma \cdot V(s')
\end{equation}
with discount factor $\gamma = 0.9$ and horizon $d_{\max} = 4$. The state value is $V(s) = \max_{a \in \mathcal{A}(s)} Q(s, a)$. For actions with multiple possible outcomes (e.g.\ credential reuse attempting several host/credential pairs), AEGIS takes the maximum over outcome values, treating the confidence score as the probability that at least one outcome succeeds.

We omit the standard failure term $(1-p) \cdot \gamma \cdot V(s)$ for two reasons. Computationally, including $V(s)$ in the recursion for $Q(s,a)$ creates self-reference requiring iterative fixed-point computation. Semantically, a failed exploit leaves state unchanged, so the failed node offers no options besides its parent.

As described in Section~\ref{sec:attack-path-simulator}, failed exploits are assigned zero value to prevent same-state retry. More precisely, if the new state equals the parent state ($s' = s_{\text{parent}}$), the branch is assigned zero value; attempts from different network positions remain viable.

\begin{table*}[b]
	\centering
	\small
	\begin{tabularx}{\textwidth}{l l X}
		\toprule
		\textbf{Factor} &        & \textbf{Question}                                                                                                   \\
		\midrule
		Per. Learning     & L1     & This scenario improved my understanding of how to respond to this type of threat.                                   \\
		\noalign{\smallskip}
		                & L2     & This scenario made me more confident in my ability to perform my role during a real incident.                       \\
		\noalign{\smallskip}
		\midrule
		Engagement      & E1 (A) & It would have been difficult to distract me when I was analysing the artefacts from this scenario.                  \\
		\noalign{\smallskip}
		                & E2 (C) & I was interested in what the adversary would do next in this scenario.                                              \\
		\noalign{\smallskip}
		                & E3 (I) & Analysing this scenario's artefacts was fun.                                                                        \\
		\midrule
		Believability   & B1     & During the exercise, I believed that this adversary's overall strategy felt real.                                   \\
		\noalign{\smallskip}
		                & B2     & During the exercise, I believed that this adversary's sequence of actions was logical and built on previous steps.  \\
		\noalign{\smallskip}
		                & B3     & During the exercise, I believed that this attack represented a credible threat to the organisation in the scenario. \\
		\midrule
		Challenge       & C1 (C) & Analysing this scenario's artefacts required me to think about many things at once.                                 \\
		\noalign{\smallskip}
		                & C2 (E) & I felt pressured and stressed while analysing the artefacts from this scenario.                                     \\
		\noalign{\smallskip}
		                & C3 (D) & The artefacts from this scenario made me think hard about prioritising what to try in the limited exercise time.    \\
		\noalign{\smallskip}
		                & C4 (P) & Analysing the artefacts from this scenario challenged my hands-on technical skills.                                 \\
		\bottomrule
	\end{tabularx}
	\caption{Questionnaire Items}
	\label{tab:questions}
\end{table*}

\subsection{Search Dynamics}

Selection uses UCT with exploration weight $C = 0.3$, favouring exploitation given expensive actions. Search terminates when the goal state is reached, no branches have positive value, or the operator intervenes. In practice, manual termination is common: operators assess generated paths and stop when a suitable candidate emerges.

With search depths of 2--4 actions, asymptotic convergence guarantees have limited relevance. The system relies on informed initialisation (Section~\ref{app:mcts:value}) rather than learning through extensive exploration.

\section{Questionnaire Design}
\label{app:questionnaire}

\subsection{Training Effectiveness Dimensions}

CDXs aim to develop mental models that prepare defenders for real incidents~\cite{gaba_simulation-based_2001, cannon-bowers_reflections_2001}. Unlike classroom learning where factual knowledge is the goal, simulation-based training develops transferable cognitive structures through experiential engagement. We organise the four dimensions by what the scenario must achieve (tranferrable mental models, indicated by perceived learning) and what it must provide to enable that outcome. Two scenario properties drive engagement: psychological fidelity (believability) and optimal challenge~\cite{maran_low-_2003, csikszentmihalyi_beyond_1975}. These affect learning through engagement, the mechanism linking instructional design to outcomes~\cite{chernikova_simulation-based_2020}.

\textbf{Perceived learning} is the primary outcome: participants' metacognitive awareness of mental model development and transfer readiness~\cite{fanning_role_2007}. Perceived learning predicts transfer intentions and real-world application~\cite{bandura_functional_2012, grossman_transfer_2011}, serving as the appropriate proximal outcome for simulation training. Unlike classroom settings where perceived and actual learning can diverge~\cite{deslauriers_measuring_2019}, simulation with debriefing enhances metacognitive accuracy through concrete performance feedback~\cite{fanning_role_2007}. Two items assess understanding improvement (L1) and role performance confidence (L2).

\textbf{Engagement} links scenario properties to learning outcomes~\cite{chernikova_simulation-based_2020}: ``fidelity matters for learning insofar as it increases engagement''~\cite{hamstra_reconsidering_2014}. Three items adapted from Webster and Hackley's~\cite{webster_teaching_1997} cognitive engagement scale measure attention focus (E1), curiosity (E2), and intrinsic interest (E3).

\textbf{Believability} operationalises psychological fidelity: the degree to which the scenario maintains the ``fiction contract,'' participants' investment in treating the simulation as real~\cite{dieckmann_deepening_2007, rudolph_establishing_2014}. If participants perceive the scenario as unrealistic, this investment suffers~\cite{saus_perceived_2010}, compromising engagement. Psychological fidelity matters more than engineering fidelity: effective learning occurs with low-fidelity simulators that capture the psychological demands of real tasks~\cite{maran_low-_2003, norman_minimal_2012}, so attack paths need not be operationally indistinguishable from real APTs. Three items assess whether the adversary's strategy felt real (B1), actions were logical (B2), and the attack represented a credible threat (B3).

\textbf{Challenge} operationalises optimal difficulty: cognitive demands appropriate to participants' skill level. Challenge affects perceived learning through two pathways: indirectly through engagement, since flow theory holds that engagement occurs when challenges match skill~\cite{csikszentmihalyi_beyond_1975, hamari_challenging_2016}; and directly through desirable difficulties, since appropriately challenging conditions enhance retention by increasing germane cognitive load~\cite{bjork_self-regulated_2013, sweller_cognitive_2011}. Because challenge contributes independently of realism, a perfectly realistic APT scenario may be counterproductive if too stealthy to leave sufficient artefacts for analysis. Four items adapted from Denisova et al.'s~\cite{denisova_measuring_2020} CORGIS scale measure cognitive (C1), emotional (C2), decision-making (C3), and performative challenge (C4).

\section{Exploratory Factor Analysis}
\label{app:efa}

We conducted exploratory factor analysis (EFA) to empirically verify the hypothesised factor structure before confirmatory analysis. We used minimum residual (MinRes) extraction with polychoric correlations (appropriate for ordinal data) and Promax oblique rotation (allowing correlated factors).

Sampling adequacy was excellent: Kaiser-Meyer-Olkin (KMO) $= 0.938$ and Bartlett's test $\chi^2 = 2765.73$, $p < 0.001$. Parallel analysis suggested 3 factors; however, we retained 4 because of (1) strong theoretical rationale from the simulation effectiveness model, (2) superior CFA fit, and (3) cross-validation stability (Tucker's $\phi > 0.99$ across random splits). Parallel analysis can under-extract when factors are highly correlated~\cite{goretzko_one_2020}. The four-factor solution explained 78.4\% of total variance (Factor 1: 30.3\%, Factor 2: 21.1\%, Factor 3: 16.1\%, Factor 4: 10.9\%).

All 12 items loaded highest on their hypothesised factors; 11 exceeded the 0.4 threshold, with L1 at 0.378 reflecting shared variance with engagement (Table~\ref{tab:efa_pattern}). Factor correlations were high (mean $|r| = 0.714$, range 0.630--0.816), consistent with a hierarchical structure confirmed via bifactor analysis (Appendix~\ref{app:cfa}).

\begin{table}[t]
	\centering
	\small
	\begin{tabularx}{\columnwidth}{c|XXXX|c}
		\toprule
		\textbf{Item} & \textbf{L}     & \textbf{E}     & \textbf{B}     & \textbf{C}     & \textbf{$h^2$} \\
		\midrule
		L1            & \textbf{0.378} &                &                &                & 0.889          \\
		L2            & \textbf{0.850} &                &                &                & 0.580          \\
		\midrule
		E1            &                & \textbf{0.678} &                &                & 0.785          \\
		E2            &                & \textbf{0.625} & 0.461          &                & 0.867          \\
		E3            &                & \textbf{0.762} &                &                & 0.765          \\
		\midrule
		B1            &                &                & \textbf{0.955} &                & 0.877          \\
		B2            &                &                & \textbf{0.813} &                & 0.840          \\
		B3            &                &                & \textbf{0.819} &                & 0.818          \\
		\midrule
		C1            &                &                &                & \textbf{0.678} & 0.776          \\
		C2            &                &                &                & \textbf{0.856} & 0.702          \\
		C3            &                &                &                & \textbf{0.561} & 0.783          \\
		C4            &                &                & 0.320          & \textbf{0.523} & 0.728          \\
		\bottomrule
	\end{tabularx}
	\caption{EFA pattern matrix. Primary loadings in bold; loadings below 0.30 suppressed.}
	\label{tab:efa_pattern}
\end{table}

\section{Confirmatory Factor Analysis \& Mediation}
\label{app:cfa}

\subsection{Factor Loadings}

Standardised factor loadings from CFA ranged from 0.686 to 0.957, all statistically significant ($p < 0.001$) (Table~\ref{tab:cfa_loadings}). Factor intercorrelations were high ($r = 0.834$--$0.903$), reflecting a hierarchical structure typical of educational measurement~\cite{chen_comparison_2006}: factors share variance through a general ``scenario quality'' dimension while retaining specific unique variance. Bifactor analysis confirmed this interpretation ($\omega_H = 0.875$), with specific factors contributing 15\% unique variance.

\begin{table}[t]
\centering
\small
\begin{tabularx}{\columnwidth}{Xc|ccc}
	\toprule
	\textbf{Factor}                & \textbf{Item} & \textbf{$\lambda$} & \textbf{SE} & \textbf{$p$} \\
	\midrule
	\multirow{2}{*}{Perceived Learning}   & L1            & 0.957              & 0.019       & $<0.001$     \\
	                               & L2            & 0.786              & 0.033       & $<0.001$     \\
	\midrule
	\multirow{4}{*}{Challenge}     & C1            & 0.887              & 0.019       & $<0.001$     \\
	                               & C2            & 0.740              & 0.035       & $<0.001$     \\
	                               & C3            & 0.926              & 0.016       & $<0.001$     \\
	                               & C4            & 0.837              & 0.026       & $<0.001$     \\
	\midrule
	\multirow{3}{*}{Engagement}    & E1            & 0.686              & 0.047       & $<0.001$     \\
	                               & E2            & 0.950              & 0.017       & $<0.001$     \\
	                               & E3            & 0.869              & 0.026       & $<0.001$     \\
	\midrule
	\multirow{3}{*}{Believability} & B1            & 0.936              & 0.014       & $<0.001$     \\
	                               & B2            & 0.943              & 0.013       & $<0.001$     \\
	                               & B3            & 0.883              & 0.022       & $<0.001$     \\
	\bottomrule
\end{tabularx}
\caption{Standardised CFA factor loadings.}
\label{tab:cfa_loadings}
\end{table}

\subsection{Discriminant Validity}

We assessed discriminant validity using HTMT2~\cite{roemer_htmt2improved_2021}, appropriate for congeneric measurement models. All HTMT2 values were below the 0.90 threshold for conceptually related constructs (range: 0.824--0.900), with bootstrap confidence intervals (5,000 replicates) supporting discriminant validity.

\subsection{Mediation Analysis}

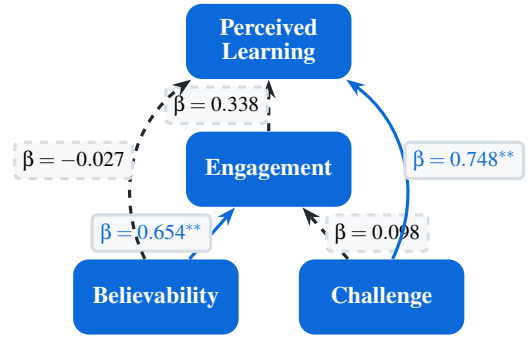
\begin{figure}[H]
\centering
\begin{tikzpicture}[
	mainnode/.style={
			rectangle,
			rounded corners=6pt,
			minimum width=2.2cm,
			minimum height=1cm,
			text centered,
			font=\rmfamily\bfseries\small,
			text=white,
			fill=blue
		},
	arrow/.style={
	-{Stealth[length=2.5mm, width=1.8mm]},
	line width=1.2pt,
	color=blue
	},
	dashedarrow/.style={
	-{Stealth[length=2.5mm, width=1.8mm]},
	line width=1.2pt,
	color=offblack,
	dashed
	},
	coef/.style={
			font=\rmfamily\footnotesize,
			fill=slate,
			draw=slateaccent,
			text=black,
			rounded corners=2pt,
			inner sep=3pt
		}
	]

	\node[mainnode] (learning) at (1.5, 3.4) {\begin{tabular}{c}Perceived\\[-2pt]Learning\end{tabular}};

	\node[mainnode] (engagement) at (1.5, 1.7) {Engagement};

	\node[mainnode] (believability) at (0, 0) {Believability};
	\node[mainnode] (challenge) at (3, 0) {Challenge};

	\draw[arrow] (believability) -- node[coef, left, pos=0.5, text=blue] {$\beta = 0.654^{**}$} (engagement);
	\draw[dashedarrow] (engagement) -- node[coef, left, pos=0.5] {$\beta = 0.338$} (learning);
	\draw[dashedarrow] (challenge) -- node[coef, right, pos=0.5] {$\beta = 0.098$} (engagement);
	\draw[arrow] (challenge) to[bend right=40] node[coef, right, pos=0.5, text=blue] {$\beta = 0.748^{**}$} (learning);
	\draw[dashedarrow] (believability) to[bend left=40] node[coef, left, pos=0.5] {$\beta = -0.027$} (learning);

\end{tikzpicture}
\caption{\label{fig:mediation} Mediation model with cluster-robust standard errors. $^{**}p < 0.01$; dashed: n.s.}
\end{figure}

We conducted exploratory mediation analysis to identify potential mechanisms, using structural equation modelling with cluster-robust standard errors to account for within-subjects dependencies ($\text{ICC} = 0.76$--$0.83$)~\cite{montoya_two-condition_2017} (Figure~\ref{fig:mediation}). These analyses are correlational, not causal; cross-sectional data prevent establishing temporal precedence~\cite{maxwell_bias_2007}, so we present these as hypothesis-generating findings.

\begin{description}
	\item[Challenge $\rightarrow$ Perceived Learning] Strong direct association ($\beta = 0.748$, $p = 0.001$), not statistically mediated through engagement.
	\item[Believability $\rightarrow$ Engagement] Strong association ($\beta = 0.654$, $p = 0.001$).
	\item[No Optimal Challenge] Modern flow theory predicts linear challenge-engagement relationships for high-importance professional activities~\cite{engeser_flow_2008}. Our data support this: the quadratic term was non-significant ($p = 0.509$), with challenge showing a linear positive association with engagement.
\end{description}

This suggests that in professional cybersecurity training, challenge is directly associated with perceived learning, while believability operates through engagement. The linear relationships supports theoretical predictions for high-stakes professional activities in literature~\cite{engeser_flow_2008}.

\subsection{Additional Methodological Notes}

\textbf{Ceiling effects.} Mean responses of 5.0--5.7 on the 7-point scale, with 11 of 12 items showing $>$15\% responses at maximum, may attenuate detected effects. Polychoric correlations and DWLS estimation address non-normality~\cite{flora_empirical_2004}.

\textbf{Path confounding.} Teams 1--3 and 4--6 received different AEGIS paths, potentially confounding path and team differences. Supplementary analysis found no significant interaction ($p > 0.05$), though power for this test was limited.

\textbf{Unbalanced design.} The 2:1 ratio of human-authored to AEGIS ratings (150:75) is handled appropriately by mixed-effects models; equivalence conclusions are robust to this imbalance.

\textbf{Statistical power.} High ICCs (0.76--0.83) yield a design effect of $\text{DEFF} \approx 2.6$ and effective sample size $N_{\text{eff}} \approx 87$, providing $>$90\% power to detect medium effects ($d = 0.50$). Multiplicity corrections across H1--H4 are omitted as each hypothesis addresses a distinct construct; all tests yielded $p < 0.001$, robust to Bonferroni correction ($\alpha_{\text{adj}} = 0.0125$).

\begin{table}[H]
\centering
\small
\begin{tabularx}{\columnwidth}{X|cccc}
	\toprule
	                  & \textbf{L} & \textbf{C} & \textbf{E} & \textbf{B} \\
	\midrule
	Perceived Learning (L)   & 1.000      &            &            &            \\
	Engagement (E)    & 0.854      & 1.000      &            &            \\
	Believability (B) & 0.839      & 0.880      & 1.000      &            \\
	Challenge (C)     & 0.903      & 0.834      & 0.902      & 1.000      \\
	\bottomrule
\end{tabularx}
\caption{Factor intercorrelations.}
\label{tab:factor_corr}
\end{table}

\begin{table}[H]
\centering
\small
\begin{tabularx}{\columnwidth}{X|cc|cc}
	\toprule
	                & \multicolumn{2}{c|}{\textbf{Human}} & \multicolumn{2}{c}{\textbf{AEGIS}}               \\
	\textbf{Factor} & $M$                                    & $SD$                               & $M$  & $SD$ \\
	\midrule
	Perceived Learning     & 5.51                                   & 1.03                               & 5.48 & 1.04 \\
	Engagement      & 5.40                                   & 1.01                               & 5.44 & 0.97 \\
	Believability   & 5.69                                   & 0.97                               & 5.66 & 0.98 \\
	Challenge       & 5.51                                   & 0.99                               & 5.50 & 0.95 \\
	\bottomrule
\end{tabularx}
\caption{Descriptive statistics by condition.}
\label{tab:descriptives}
\end{table}

\end{document}